\newcommand{\be}{\begin{equation}}
\newcommand{\ee}{\end{equation}}
\def\bea{\begin{eqnarray}}
\def\eea{\end{eqnarray}}

\documentstyle[sprocl]{article}
\title{Neutrino Masses  \\and\\Lepton-Quark Symmetries}

\author{P. Ramond}
\address{Institute for Fundamental Theory\\
Department of Physics, University of Florida\\ 
Gainesville, Fl 32611}

\begin{document}

%\begin{abstract}
%\end{abstract}
% typeset front matter (including abstract)

\maketitle\abstracts{
The recent discovery by SuperKamiokande of evidence for neutrino
masses requires the addition of at least seven new parameters to the Standard 
Model. We discuss the general theoretical schemes which require their 
inclusion, and point out how quark-lepton symmetries, either in the 
framework of Grand Unification, or of string theories, can be used to determine them}

\section{Neutrino Story}
%\section{Grand Unification}
The Neutrino Story starts with the experiment of 
O. Von Bayer, O. Hahn, and L. Meitner~\cite{RAD,VOL} who  measured the
spectrum of 
electrons in $\beta$ radioactivity, and found it to be discrete! 
In 1914, Chadwick~\cite{CHAD}, then 
 in Geiger's laboratory in Berlin 
came to the correct conclusion of a continuous 
electron spectrum, but was interned for the duration of the Great War. 
After much controversy, the issue was settled in 1927 by  C.D. 
Ellis and W. A. Wooster~\cite{ELWO}, who  found the mean energy 
liberated in $\beta$ 
decay accounted to be only $1/3$ of the allowed 
energy. The stage was set for  W. Pauli's famous 1930 letter. 

In December of that year, in a  letter that starts with typical panache, $``${\it  
Dear Radioactive Ladies and Gentlemen...}", W. Pauli puts forward 
a ``{\it desperate}" way out: there is a companion particle 
to the $\beta$ electron. Undetected, it must be electrically 
neutral, and in order to balance the $N-Li^6$ statistics, it 
carries spin $1/2$. He calls it the {\it neutron}, but sees no 
reason why it could not be massive. 

In 1933, E. Fermi in his  $\beta$ decay paper 
gave it its final name, the little neutron or {\it neutrino}, as it is
clearly much lighter than Chadwick's neutron which had just been discovered.
 
In 1945,  B. Pontecorvo~\cite{PONTA}
proposes the unthinkable: neutrinos can be detected, through 
the following
observation: an electron neutrino that hits a ${^{37}Cl}$ atom will
transform it into the inert radioactive gas ${^{37}Ar}$, which then can be stored and be  detected through its radioactive decay.  Pontecorvo did not publish
the report, perhaps because of its secret classification, or because Fermi  thought the idea ingenious but not immediately  achievable. 
In 1954,  Davis~\cite{DAVISA} follows up on Pontecorvo's original
proposal, by setting a tank of cleaning fluid outside a nuclear reactor. 

In 1956, using a scintillation counter experiment they had proposed three years earlier~\cite{COREA},  Cowan and Reines~\cite{COREB} discover electron antineutrinos 
 through the reaction
$\overline \nu_e+p\rightarrow e^+ +n$. Cowan passed away before 1995,
the year Fred Reines was awarded the Nobel Prize for their discovery. 
There emerge two  lessons in neutrino physics: not only is patience
required but also longevity: it took $26$ years from birth to
detection and then another $39$ for the Nobel Committee to recognize 
the achievement! This should encourage future  physicists to 
train their children at the earliest age to follow their footsteps, 
in order to establish dynasties of neutrino physicists.  

In 1956, it was rumored that Davis had found evidence for
neutrinos coming from a pile, and Pontecorvo~\cite{PONTB}, influenced by the recent work of   Gell-Mann and Pais, theorized that  an
 antineutrino produced in the Savannah reactor could oscillate
into a neutrino and be detected by Davis. The rumor went
away, but the idea of neutrino oscillations was born; it has remained
with us ever since, and proven the most potent tool in hunting for
neutrino masses. 

Having detected neutrinos, there remained to determine its spin and
mass. Its helicity was measured in 1958 by M. Goldhaber~\cite{MGOLD},
but convincing evidence  for its mass had, until SuperK's bombshell, eluded
experimentalists. 

After the 1957 Lee and Yang proposal of parity  violation, the neutrino is again at the center of the action. Unlike the charged elementary particles which have both left- and right-handed components, weakly interacting neutrinos are purely left-handed (a
ntineutrinos are right-handed), which means that lepton-number is chiral. 

The second neutrino, the muon neutrino  is detected~\cite{2NEUT} in 1962, (long anticipated by theorists Inou\"e and Sakata in 1943~\cite{INSA}). This time things went a bit faster as it took only 19 years from theory (1943) to discovery (1962) and 26 yea
rs to Nobel recognition (1988). 

That same year, Maki, Nakagawa and Sakata~\cite{MANASA} introduce two crucial ideas; one is that these two neutrinos can mix, and the second is that this mixing can cause one type of neutrino to oscillate into the other (called today flavor oscillation). 
This is possible only if the two neutrino flavors have different masses.

In 1963, the Astrophysics group at Caltech, Bahcall, Fowler,
Iben and Sears~\cite{BWIS} puts forward the most accurate of neutrino
fluxes from the Sun. Their calculations included the all important
Boron decay spectrum, which produces neutrinos with the right energy
range for the Chlorine experiment. 

In 1964, using Bahcall's result~\cite{BAH} of an enhanced capture rate of ${^8B}$ neutrinos through an excited state of ${^{37}Ar}$, Davis~\cite{DAVISB}
proposes to search for ${^8B}$ solar neutrinos using a $100,000$ gallon
tank of cleaning fluid deep underground.  Soon after, R. Davis starts 
his epochal experiment at the Homestake mine, marking the
beginning of the solar neutrino watch which continues to this day. In
1968, Davis et al reported~\cite{DAVISC} a deficit in the solar neutrino flux, a result that has withstood scrutiny to this day, and stands as a truly
remarkable experimental {\it tour de force}. Shortly after, Gribov and
Pontecorvo~\cite{GRIPO} interpreted the deficit as evidence for neutrino oscillations.

In the early 1970's, with the idea of quark-lepton symmetries~\cite{PASA,GG} comes the idea that the proton could be unstable. This brings about the construction of  underground (to avoid contamination from cosmic ray by-product) detectors, large enough t
o monitor many protons, and instrumentalized to detect the \v Cerenkov light emitted by its decay products. By the middle 1980's, several such detectors are in place. They fail to detect proton decay, but in a serendipitous turn of events, 150,000 years e
arlier, a supernova erupted in the large Magellanic Cloud, and in 1987, its burst of neutrinos was detected in these detectors! All of a sudden, proton decay detectors turn their attention to neutrinos, and to this day still waiting for its protons to dec
ay!

As we all know, these detectors routinely monitor neutrinos from the Sun, as well as neutrinos produced by cosmic ray collisions. 

\section{Standard Model Neutrinos}
The standard model of electro-weak and strong interactions contains three left-handed neutrinos.  The three neutrinos are represented by two-components Weyl spinors, $\nu^{}_{i}$, $i=e,\mu,\tau$, each describing a left-handed fermion (right-handed antifer
mion). As the upper components of weak isodoublets $L^{}_i$, they have $I^{}_{3W}=1/2$, and a unit of the global $i$th lepton number. 

These standard model neutrinos are strictly massless. The only Lorentz scalar made out of these neutrinos is the Majorana mass, of the form
$\nu^{t}_{i}\nu^{}_{j}$; it has the quantum numbers of a weak isotriplet, with third component  $I^{}_{3W}=1$, as well as two units of total lepton number.  Higgs isotriplet with two units 
 of lepton number could generate neutrino Majorana masses, but there is no such higgs in the Standard Model:  there are no tree-level neutrino masses in the standard model.

Quantum corrections, however,  are  not limited to
renormalizable couplings, and it is easy to make a weak isotriplet out
of two isodoublets, yielding the $SU(2)\times U(1)$ invariant
$L^t_i\vec\tau L^{}_j\cdot H^t_{}\vec\tau H$, where $H$ is the Higgs
doublet. As  this term is not invariant under lepton number, it is not be generated in perturbation theory. Thus the
important conclusion: {\it The standard model neutrinos are kept
massless by global chiral lepton number symmetry}. The detection of non-zero neutrino masses is therefore {\it a tangible indication of physics beyond the standard model}.

\section{Neutrino Mass Models}
Neutrinos must be extraordinarily light: experiments indicate  $m_{\nu_e}< 
10~ {\rm eV}$, $m_{\nu_\mu}< 170~ {\rm keV}$, $m_{\nu_\tau}< 18~ {\rm MeV}$~\cite{PDG}, and any model of neutrino masses must explain this suppression.

We do not discuss generating neutrino masses without new fermions, by  
breaking  lepton number through interaction of lepton number-carrying Higgs fields.

The  natural way to generate neutrinos masses is to introduce for
each one its electroweak singlet Dirac partner, $\overline
N^{}_i$. These appear naturally in the Grand Unified group
$SO(10)$ where they complete each family into its spinor representation. Neutrino Dirac masses stem from the couplings $L^{}_i\overline 
N^{}_j H$ after electroweak breaking. Unfortunately, these Yukawa
couplings yield masses which are too big, 
of the same order of magnitude as the masses of the charged
elementary particles $m\sim\Delta I_w=1/2$. 

The situation is remedied by introducing
Majorana mass terms $\overline N^{}_i\overline N^{}_j$ for the
right-handed neutrinos. The masses of these new degrees of freedom are 
arbitrary, as they have no electroweak quantum numbers, $M\sim\Delta
I_w=0$. If they are much larger than the electroweak scale, the
neutrino masses are suppressed relative to that of their charged
counterparts by the ratio of the electroweak scale to that new scale: 
the mass matrix  (in $3\times 3$ block form) is
\be
\hskip 1in \pmatrix{0& m\cr m&M}\ ,
\ee
leading, for each family,  to one small and one large eigenvalue 
\be
m_\nu~\sim~ m\cdot {m\over M}~\sim~ \left(\Delta I_w={1\over 2}\right)\cdot 
\left({ \Delta I_w={1\over 2}
\over \Delta I_w=0 }\right)\ .\ee 
This seesaw mechanism~\cite{SEESAW} provides a natural
explanation for  small neutrino masses as long as lepton
number is broken at a large scale $M$. With $M$ around the energy at
which the gauge couplings unify, this yields neutrino masses at or
below tenths of eVs, consistent with the SuperK results. 

The lepton flavor mixing comes from  the diagonalization
of the charged lepton Yukawa couplings, and  of the neutrino
mass matrix. From the charged lepton Yukawas, we obtain ${\cal U}_e^{}$, 
the unitary matrix that rotates the
lepton doublets $L^{}_i$. From the neutrino Majorana matrix, we obtain
$\cal U_\nu$, the matrix that diagonalizes the Majorana mass matrix. 
The $6\times 6$ seesaw Majorana matrix can be written in $3\times 3$
block form
\be
{\cal M}={\cal V}_\nu^t ~{\cal D} {\cal V}^{}_\nu\sim\pmatrix {{\cal
U}_{\nu\nu}&\epsilon {\cal U}^{}_{\nu N}\cr
\epsilon{\cal U}^{t}_{N \nu}&{\cal U}^{}_{NN}\cr}\ ,\ee
where $\epsilon$ is the tiny ratio of the electroweak to lepton
number violating scales, and ${\cal D}={\rm diag}(\epsilon^2{\cal D}_\nu, {\cal D}_N)$,
 is a diagonal matrix. ${\cal D}_\nu$ contains the
three neutrino masses, and $\epsilon^2$ is the seesaw suppression. The
weak charged current is then given by
\be
j^+_\mu=e^\dagger_i\sigma_\mu {\cal U}^{ij}_{MNS}\nu_j\ ,\ee
where
\be
{\cal U}^{}_{MNS}={\cal U}^{}_e{\cal U}^\dagger_\nu\ ,\ee
is the Maki-Nakagawa-Sakata~\cite{MANASA} (MNS) flavor mixing matrix, the analog of the CKM matrix in the quark sector. 

In the seesaw-augmented standard model, this mixing matrix is totally 
arbitrary. It contains, as does the CKM matrix, three rotation angles,
and one CP-violating phase. In the seesaw scenario, it also contains
 two additional CP-violating phases
which cannot be absorbed in a redefinition of the neutrino
fields, because of their Majorana masses (these extra phases can be
measured only in $\Delta {\cal L}=2$ processes). These  additional
parameters of the seesaw-augmented standard model, need to  be determined by
experiment.
\section{Theories}
Theoretical predictions of lepton hierarchies and
mixings depend very much on hitherto untested theoretical assumptions. In the quark sector, where the bulk of the experimental data resides,
the theoretical origin of quark hierarchies and mixings is a mystery,
although there exits many theories, but none so convincing as to offer a
definitive answer to the community's satisfaction. It is therefore no surprise that there are  more theories of lepton
masses and mixings than there are parameters to be measured. Nevertheless, one can 
present the issues as questions:
\begin{itemize}
\item Do the right handed neutrinos have quantum numbers beyond the
standard model?
\item Are quarks and leptons related by grand unified theories?
\item Are quarks and leptons related by anomalies?
\item Are there family symmetries for quarks and leptons?
\end{itemize}

The measured numerical value of
the neutrino mass difference (barring any fortuitous degeneracies), suggests 
through the seesaw mechanism, a mass for the right-handed neutrinos
that is consistent with the scale at which
the gauge couplings unify. Is this just a numerical  coincidence, 
or should we view this as a hint for grand unification?

Grand unified Theories, originally proposed as a way to treat
leptons and quarks on the same footing,  imply  symmetries much larger than 
the standard model's. Implementation of these
ideas necessitates a desert and supersymmetry, but also a carefully designed
contingent of Higgs particles to achieve the desired symmetry
breaking. That such models can be built is perhaps more of a testimony
to the cleverness of theorists rather than of Nature's. Indeed with the
advent of string theory, we know that the best features of grand
unified theories can be preserved, as most of the symmetry breaking is
achieved by geometric compactification from higher dimensions~\cite{CANDELAS}.

An alternative point of view is that the vanishing of 
chiral anomalies is necessary for consistent  theories,
and their cancellation is most easily achieved by assembling matter in
representations of anomaly-free groups. Perhaps anomaly cancellation
is more important than group structure.

Below, we present two theoretical frameworks of our work, in which one deduces
the lepton mixing parameters and masses. One is ancient~\cite{HRR}, uses the standard techniques of grand unification, but it had the virtue of {\it predicting} the large
$\nu_\mu-\nu_\tau$ mixing observed by SuperKamiokande. The other~\cite{ILR} is more recent, and uses
extra Abelian family symmetries to explain both quark and lepton
hierarchies. It also predicts large $\nu_\mu-\nu_\tau$ mixing. Both
schemes imply small $\nu_e-\nu_\mu$ mixings.

\subsection{A Grand Unified Model}  
The seesaw mechanism was born in the context of the grand unified group $SO(10)$, which
naturally contains electroweak neutral right-handed neutrinos. Each standard model family is contained in 
two irreducible representations of $SU(5)$. However, the predictions of this
theory for Yukawa couplings is not so clear cut, and to reproduce the
known quark and charged lepton hierarchies, a special but simple set of Higgs
particles had to be included. In the simple scheme proposed by Georgi
and Jarlskog~\cite{GJ}, the ratios between the charged leptons and quark masses
is reproduced, albeit not naturally since two Yukawa couplings, not
fixed by group theory, had to be set equal. This motivated us to
generalize their scheme to $SO(10)$, where their scheme was
(technically) natural, which meant that we had an automatic window
into neutrino masses through the seesaw. The Yukawa couplings were of
the form

\be
[A{\bf 16}_1\cdot {\bf 16}_2+B{\bf 16}_3{\bf 16}_3]\cdot{\bf 126}_1
+[a{\bf 16}_1\cdot {\bf 16}_2+b{\bf 16}_3{\bf 16}_3]\cdot({\bf
10}_1+i{\bf 10}_2)\ee
\be
+c{\bf 16}_2{\bf 16}_2\cdot{\overline{\bf 126}}_2+d{\bf 16}_2{\bf 16}_3
{\overline{\bf 126}}_3\ .
\ee
This is of course Higgs-heavy, but the attitude at the time 
was ``damn the Higgs torpedoes, and see what happens". This assignment was ``technically'' natural, enforced by two discrete
symmetries. A modern treatment would include non-renormalizable
operators~\cite{BPW}, rather than introducing the ${\bf 126}$ representations, which spoil asymptotic freedom.

The Higgs  vacuum values produced  the resultant masses 
\be m_b=m_\tau\ ;\qquad m_dm_s=m_em_\mu\ ;\qquad m_d-m_s=3(m_e-m_\mu)\
.\ee
and mixing angles
\be
V_{us}=\tan\theta_c=\sqrt{m_d\over m_s}\ ;\qquad V_{cb}=\sqrt{m_c\over
m_t}\ .\ee
While reproducing the well-known lepton and quark mass hierarchies, it
predicted a long-lived $b$ quark, contrary to the lore of the time.
It also made predictions in the lepton sector, namely
{\bf maximal} $\nu_\tau-\nu_\mu$ mixing, small $\nu_e-\nu_\mu$
mixing of the order of $(m_e/m_\mu)^{1/2}$, and no $\nu_e-\nu_\tau$
mixing. 

The neutral lepton masses came out to be hierarchical, but heavily
dependent on the masses of the right-handed neutrinos. The
 electron neutrino mass came out much lighter
than those of $\nu_\mu$ and $\nu_\tau$. Their numerical values
depended on the top quark mass, which was then supposed to be in the
tens of GeVs!

Given the present knowledge, some of the features are remarkable, such as
the long-lived $b$ quark and the maximal $\nu_\tau-\nu_\mu$
mixing. On the other hand, the actual numerical value of the $b$ lifetime was off
a bit,and the $\nu_e-\nu_\mu$ mixing was too large to reproduce the small angle
MSW solution of the solar neutrino problem. 

The lesson should be that the simplest $SO(10)$ model 
that fits the observed quark and charged lepton hierarchies, reproduces, at least qualitatively, the maximal mixing found by
SuperK, and predicts small mixing with the electron neutrino~\cite{CASE}.

\subsection{A Non-grand-unified  Model}
There is another way to generate hierarchies, based on adding extra
family symmetries to the standard model, without invoking grand
unification. These types of models address only the Cabibbo
suppression of the Yukawa couplings, and are not as predictive as
specific grand unified models. Still, they predict no Cabibbo
suppression between the muon and tau neutrinos. Below, we present a
pre-SuperK  model~\cite{ILR} with those features. 

The Cabibbo supression is assumed to be an indication of extra
family symmetries in the standard model. The idea is that any standard model-invariant
operator, such as ${\bf Q}_i{\bf \overline d}_jH_d$,  cannot be present
at tree-level if there are additional symmetries under which the
operator is not invariant. Simplest is to assume an Abelian symmetry,
with an electroweak singlet field $\theta$,  as its order parameter.
Then  the interaction
\be
{\bf Q}_i{\bf \overline d}_jH_d\left({\theta\over M}\right)^{n_{ij}}\ee
can appear in the potential as long as the family charges balance under the
new symmetry. As $\theta$ acquires a $vev$, this leads to a
suppression of the Yukawa couplings of the order of $\lambda^{n_{ij}}$
for each matrix element, with 
$\lambda=\theta/M$ identified with  the Cabibbo angle, and
$M$ is the natural cut-off of the effective low energy  theory. 
As a consequence of the charge balance equation
\be X_{if}^{[d]}+n^{}_{ij}X^{}_\theta=0\ ,\ee
the exponents of the suppression are related to the charge of the
standard model-invariant operator~\cite{FN},  the sum of the
charges of the fields that make up the the invariant. 

This simple Ansatz, together with the seesaw mechanism, 
implies that the family structure of the neutrino mass matrix is
determined by the charges of the left-handed lepton doublet fields. 

Each charged lepton Yukawa coupling 
$L_i\overline N_j H_u$, has an extra  charge $X_{L_i}+X_{Nj}+X_{H}$, which
gives the Cabibbo suppression of the $ij$ matrix element. Hence, 
 the orders of magnitude of these couplings can be expressed as 
\be
\pmatrix{\lambda^{l_1}&0&0\cr
0&\lambda^{l_2}&0\cr
0&0&\lambda^{l_3}\cr}{\hat Y}\pmatrix{\lambda^{p_1}&0&0\cr
0&\lambda^{p_2}&0\cr
0&0&\lambda^{p_3}\cr}\ ,\ee
where ${\hat Y}$ is a Yukawa matrix with no Cabibbo
suppressions, $l_i=X_{L_i}/X_\theta$ are the charges of the
left-handed doublets, and 
$p_i=X_{N_i}/X_\theta$, those of the singlets. The first matrix forms half of
the MNS matrix. Similarly, the mass matrix for the right-handed
neutrinos, $\overline N_i\overline N_j$ will be written in the form
\be
\pmatrix{\lambda^{p_1}&0&0\cr
0&\lambda^{p_2}&0\cr
0&0&\lambda^{p_3}\cr}{\cal M}\pmatrix{\lambda^{p_1}&0&0\cr
0&\lambda^{p_2}&0\cr
0&0&\lambda^{p_3}\cr}\ .\ee
The diagonalization of the  seesaw matrix is of the form 
\be 
L_iH_u\overline N_j \left({1\over{{\overline N}~\overline
N}}\right)_{jk}\overline N_kH_uL_l\ ,\ee
from which the Cabibbo suppression matrix from the $\overline N_i$
fields {\it cancels}, leaving us with
\be
 \pmatrix{\lambda^{l_1}&0&0\cr
0&\lambda^{l_2}&0\cr
0&0&\lambda^{l_3}\cr}\hat{\cal M}\pmatrix{\lambda^{l_1}&0&0\cr
0&\lambda^{l_2}&0\cr
0&0&\lambda^{l_3}\cr}\ ,\ee
where $\hat{\cal M}$ is a matrix with no Cabibbo suppressions.  
The Cabibbo structure of the seesaw neutrino matrix is determined
solely by the charges of the lepton doublets! As a result, the Cabibbo
structure of the MNS
mixing matrix is also due entirely to the charges of the three lepton
doublets. This general conclusion depends on the existence of at least
one Abelian family symmetry, which we argue is implied by the observed
structure in the quark sector.

The Wolfenstein parametrization of the CKM matrix~\cite{WOLF}, 
\be
\hskip 1in
\pmatrix{1&\lambda & \lambda ^3\cr
             \lambda &1&\lambda ^2\cr 
             \lambda ^3&\lambda ^2&1}\ ,\ee
and the Cabibbo structure of the quark mass ratios
\be {m_{u}\over m_t}\sim \lambda ^8\;\;\;{m_c\over m_t}\sim 
\lambda ^4\;\;\; ;
\;\;\; {m_d\over m_b}\sim \lambda ^4\;\;\;{m_s\over m_b}\sim
\lambda^2\ ,\ee
can be reproduced~\cite{ILR,EIR} by a simple {\it family-traceless} charge assignment 
for the three quark families, namely
\be
X_{{\bf Q},{\bf \overline u},{\bf \overline d}} ={\cal B}(2,-1,-1)+
\eta_{{\bf Q},{\bf \overline u},{\bf \overline d}}(1,0,-1)\ ,\ee
where ${\cal B}$ is baryon number, $\eta_{{\bf \overline d}}=0 $, and 
$\eta_{{\bf Q}}=\eta_{{\bf \overline u}}=2$. 
Two striking facts are evident: 
\begin{itemize}
\item the charges of the down quarks, ${\bf \overline d}$, 
associated with the second and third families are the same, 
\item ${\bf Q}$ and ${\bf \overline u}$ have the same value for 
$\eta$.
\end{itemize}
To relate these quark charge assignments to those of the leptons,
we need to inject some more theoretical prejudices. Assume these
 family-traceless charges are gauged, and not anomalous. Then to
cancel anomalies, the leptons must themselves have family charges. 

Anomaly cancellation generically implies group structure. In $SO(10)$, 
baryon number generalizes to ${\cal B}-{\cal L}$, where ${\cal L}$ is total
lepton number, and  in 
$SU(5)$   the fermion assignment is ${\bf\overline 5}={\bf
\overline d}+L$, and ${\bf 10}={\bf Q}+{\bf \overline u}+\overline
e$. Thus anomaly cancellation is easily achieved by
assigning $\eta=0$ to the lepton doublet $L_i$, and $\eta=2$ to the
electron singlet $\overline e_i$, and by generalizing baryon number to
${\cal B}-{\cal L}$, leading to the charges
\be
X_{{\bf Q},{\bf \overline u},{\bf \overline d}, L,\overline e} =({\cal
B}-{\cal L})(2,-1,-1)+
\eta_{{\bf Q},{\bf \overline u},{\bf \overline d}}(1,0,-1)\ ,\ee
where now $\eta_{{\bf \overline d}}=\eta_{L}=0 $, and 
$\eta_{{\bf Q}}=\eta_{{\bf \overline u}}=\eta_{\overline e}=2$. 
It is interesting to note that $\eta$ is at least in $E_6$. 
The origin of such charges is not clear, as it implies in the
superstring context, rather unconventional compactification.

As a result,  the charges of the
lepton doublets are simply $X_{L_i}=-(2,-1,-1)$. We have just
argued that these charges determine the Cabibbo structure of the MNS
lepton mixing matrix to be
\be
\hskip .2in
{\cal U}^{}_{MNS}\sim\pmatrix{1&\lambda^3&\lambda^3\cr
\lambda^3&1&1\cr \lambda^3&1&1\cr}\ ,\ee
implying{\it  no Cabibbo suppression in the mixing between
$\nu_\mu$ and $\nu_\tau$}. This is consistent with the SuperK
discovery and  with the small angle MSW~\cite{MSW} solution to the solar
neutrino deficit. One also obtains a much lighter electron neutrino, and
Cabibbo-comparable masses for the muon and tau neutrinos. Notice that
these predictions are  subtly different from those
of grand unification, as they  yield $\nu_e-\nu_\tau$ mixing. 
It also implies a much lighter electron neutrino, and
Cabibbo-comparable masses for the muon and tau neutrinos. 

On the other hand, the scale of the neutrino mass values depend on the family 
trace of the family charge(s). Here we simply quote the results
our model~\cite{ILR}. The masses of the right-handed neutrinos are found to be of the
following orders of magnitude
\be
m_{\overline N_e}\sim M\lambda^{13}\ ;\qquad m_{\overline N_\mu}\sim
m_{\overline N_\tau}\sim M\lambda^7\ ,\ee
where $M$ is the scale of the right-handed neutrino mass terms,
assumed to be the cut-off. The seesaw mass matrix for the three light  neutrinos 
comes out to be 
\be
\hskip .5in
 m^{}_0\pmatrix{a\lambda^6&b\lambda^3&c\lambda^3\cr
b\lambda^3&d&e\cr
c\lambda^3&e&f\cr}\ ,\ee
where we have added for future reference the prefactors $a,b,c,d,e,f$, all of 
order one, and 
\be m_0^{}={v_u^2\over
{M\lambda^3}}\ ,\ee
where $v_u$ is the $vev$ of the Higgs doublet. This matrix has one light eigenvalue
\be
m_{\nu_e}\sim m_0^{}\lambda^6_{}\ .\ee
Without a detailed analysis of the prefactors, the masses of the other 
two neutrinos come out  to be both of 
 order $m_0$. 
The mass difference announced by  superK~\cite{SUPERK}  cannot  be
reproduced without going beyond the model, by taking into account the
prefactors. The two heavier mass
eigenstates and their mixing angle are written in terms of 
\be
x={df-e^2\over (d+f)^2}\ ,\qquad y={d-f\over d+f}\ ,\ee
as
\be {m_{\nu_2}\over m_{\nu_3}}={1-\sqrt{1-4x}\over 1+\sqrt{1-4x}}\
,\qquad \sin^22\theta_{\mu\tau}=1-{y^2\over 1-4x}\ .\ee
If $4x\sim 1$, the two heaviest neutrinos are nearly degenerate. If
$4x\ll 1$, a condition easy to achieve if $d$ and $f$ have the same
sign, we can obtain an adequate split between the two mass
eigenstates. For illustrative purposes, when $0.03<x<0.15$, we find
\be
4.4\times 10^{-6}\le \Delta m^2_{\nu_e-\nu_\mu}\le 10^{-5}~{rm eV}^2\
  ,\ee
which yields the correct non-adiabatic MSW effect, and
\be
5\times 10^{-4}\le  \Delta m^2_{\nu_\mu-\nu_\tau}\le 5\times
10^{-3}~{\rm eV}^2\ ,\ee
for the atmospheric neutrino effect. These were calculated with a
cut-off, $10^{16}~{\rm GeV}<M<4\times 10^{17}~{\rm GeV}$, and a mixing
angle, $0.9<\sin^22\theta_{\mu-\tau}<1$. This value of the cut-off is 
 compatible not only with the data but also with the gauge
coupling unification scale, a necessary condition for the consistency
of our model, and more generally for  the basic ideas of Grand Unification. 

\section{Outlook}
Exact predictions of neutrino masses and mixings depend on developing
a credible theory of flavor. In the absence of such, we have presented
two schemes, which predicted not only maximal $\nu_\mu-\nu_\tau$
mixing, but also smal $\nu_e-\nu_\mu$ mixings. Neither scheme includes
sterile neutrinos. The present experimental situation is somewhat
unclear: the LSND results~\cite{LSND} imply the presence of a sterile neutrino; at
this conference we heard that superK favors 
$\nu_\mu-\nu_\tau$ oscillation over $\nu_\mu-\nu_{\rm sterile}$, and the origin
of the solar neutrino deficit remains  a puzzle, which several
possible explanations. One is the non-adiabatic MSW effect in
the Sun, which our theoretical ideas seem to favor. However, it is an
experimental question which is soon to be answered by the
continuing monitoring of the $^8B$ spectrum by SuperK, and the advent  of the SNO detector. Neutrino physics is at an exciting stage, and
experimentally vibrant, as upcoming measurements will help us
determine iour basic ideas about fundamental interactions.
\section{Acknowledgments}
I would like to thank Professors G. Domokos and S. K\"ovesi-Domokos
for their usual superb hospitality and the high scientific quality of
this workshop. This research was supported in part by the department
of energy under grant DE-FG02-97ER41029.

\end{document}